# Dielectric catastrophe at the Mott and Wigner transitions in a moiré superlattice

Yanhao Tang[1,2*], Jie Gu[1], Song Liu[3], Kenji Watanabe[4], Takashi Taniguchi[4], James C. Hone[3], Kin Fai Mak[1,5,6*], and Jie Shan[1,5,6*]

[1]School of Applied and Engineering Physics, Cornell University, Ithaca, NY, USA
[2]Interdisciplinary Center for Quantum Information, Zhejiang Province Key Laboratory of Quantum Technology, and Department of Physics, Zhejiang University, Hangzhou 310027, China
[3]Department of Mechanical Engineering, Columbia University, New York, NY, USA
[4]National Institute for Materials Science, 1-1 Namiki, 305-0044 Tsukuba, Japan
[5]Laboratory of Atomic and Solid State Physics, Cornell University, Ithaca, NY, USA
[6]Kavli Institute at Cornell for Nanoscale Science, Ithaca, NY, USA

These authors contributed equally: Yanhao Tang, Jie Gu
*Email: yanhaotc@zju.edu.cn; kinfai.mak@cornell.edu; jie.shan@cornell.edu

**The metal-insulator transition (MIT) driven by electronic correlations is a fundamental and challenging problem in condensed-matter physics [1-3]. Particularly, whether such a transition can be continuous remains open. The emergence of semiconducting moiré materials with continuously tunable bandwidth provides an ideal platform to study interaction-driven MITs [4-10]. Although a bandwidth-tuned MIT at fixed full electron filling of the moiré superlattice has been reported recently [5,6], that at fractional filling, which involves translational symmetry breaking of the underlying superlattice [9, 11-14], remains elusive. Here, we demonstrate bandwidth-tuned MITs in a $MoSe_2/WS_2$ moiré superlattice at both integer and fractional fillings using the exciton sensing technique [15, 16]. The bandwidth is controlled by an out-of-plane electric field. The dielectric response is probed optically with the 2s exciton in a remote $WSe_2$ sensor layer. The exciton spectral weight is negligible for the metallic state, consistent with a large negative dielectric constant. It continuously vanishes when the transition is approached from the insulating side, corresponding to a diverging dielectric constant or a 'dielectric catastrophe' [17-20]. Our results support continuous interaction-driven MITs in a two-dimensional triangular lattice and stimulate future explorations of exotic quantum phases, such as quantum spin liquids, in their vicinities [9, 11, 12, 14, 21, 22].**

MITs accompanied by large electrical conductivity change are widely observed in condensed-matter systems [1-3, 23, 24]. One particularly intriguing origin of charge localization and insulating phases is Coulomb repulsion between electrons [25]. The prototype model for interacting electrons in a lattice is the single-band Hubbard model with electronic bandwidth, $W$, and on-site Coulomb repulsion, $U$. The ground state of the electronic system for half band or full lattice filling is a Mott insulator in the strong interaction limit ($U \gg W$), and a Fermi liquid in the weak interaction limit ($U \ll W$). A MIT, the Mott transition, is expected near $U \sim W$ (Ref. [25]). Similarly, if the electrons are localized by the extended Coulomb repulsion, $V$, a Wigner-Mott insulator that spontaneously breaks the underlying lattice space-group symmetries is formed at fractional lattice filling [15, 26-28]; a Wigner-Mott transition or Wigner transition for short is expected near $V \sim W$ (Ref. [9, 11-14]). The evolution of a Mott or Wigner-Mott insulator into a metal as a function of the interaction strength remains a challenging theoretical problem [9, 11-14, 23]; it also raises an exciting opportunity for realizing exotic quantum phases, such as spin liquids, near the transition if it is continuous [9, 11, 12, 14, 21, 22].

Experimentally, the interaction strength, or equivalently, the bandwidth can be tuned by applying isostatic or chemical pressure. In almost all known materials, the MIT is driven first order [1]. But the conclusion is complicated by the fact that these systems typically show magnetic or structural ordering on the insulating side; it is difficult for two kinds of orders to vanish simultaneously [1, 21]. The recent experimental breakthroughs in semiconducting transition metal dichalcogenide (TMD) moiré materials [15, 26, 29-31] have opened a new avenue to explore the interaction-driven MITs [4-10]. These materials form a two-dimensional (2D) triangular lattice that suppresses magnetic ordering. The electrons are trapped by the periodic moiré potential; they can tunnel between the moiré sites and interact with each other via both the on-site and extended Coulomb repulsions. This realizes the triangular lattice extended Hubbard model [4, 13, 32, 33]. A variety of correlated insulating states are reported, including the Mott insulator [26, 29-31] at odd integer filling and Wigner-Mott insulators [15, 26-28, 34] at fractional fillings of the moiré superlattice. A continuous bandwidth-tuned Mott transition is also observed by electrical measurements [5, 6]. However, the Wigner transition remains elusive.

Here we report the observation of bandwidth-tuned Mott and Wigner transitions in an angle-aligned $MoSe_2/WS_2$ heterobilayer by the exciton sensing technique [15, 16]. The heterobilayer forms a triangular moiré lattice with a lattice density of $\approx 1.9 \times 10^{12}$ cm$^{-2}$. It is encapsulated in hexagonal boron nitride (hBN) and gated by a top and bottom few-layer graphite gate (Fig. 1a). The dual-gate structure enables independent tuning of the electron density in the moiré lattice $\nu$ (in units of the moiré density), and the out-of-plane electric field, $E$ (> 0 for field pointing from the $MoSe_2$ to $WS_2$ layer). The $MoSe_2/WS_2$ heterobilayer has a type-I band alignment with both the conduction and valence band edges located in the $MoSe_2$ layer (Fig. 1b). The band offsets are $\Delta_c \sim 100$ meV for the conduction bands and $\Delta_v \sim 320$ meV for the valence bands from optical spectroscopy measurement (Methods). We tune the electronic bandwidth at a fixed doping density by the electric-field effect [5, 6]. The out-of-plane electric field varies the moiré potential depth by controlling the band offset and the resonance interlayer hopping amplitude because spatial modulation in the interlayer hopping amplitude is one origin of the periodic moiré potential [4, 5]. Because $\Delta_c < \Delta_v$, the electric-field effect on the conduction bandwidth is much larger; we focus on the case of electron doping.

The dielectric response of the moiré system to an electric field is probed by the exciton sensing technique. Recent studies show that the technique is highly sensitive to the insulating states [15, 16]. These states perturb the electric field between the optically excited electrons and holes (excitons) in a charge neutral $WSe_2$ monolayer that is separated from the moiré superlattice by a bilayer hBN. The spacer thickness is smaller than the Bohr radius of the 2s and higher-energy exciton states. We probe the effective dielectric constant, $\varepsilon$, of the moiré heterobilayer by measuring the 2s exciton that has the largest spectral weight. Both the 2s exciton resonance energy and spectral weight, $S$, depend on $\varepsilon$ (Ref. [35]). We analyze the spectral weight near the MITs (Methods); it is difficult to determine the exciton binding energy accurately. Unless otherwise specified, all measurements are performed at 3.6 K; the corresponding thermal excitation energy is substantially below the characteristic energy scales ($U$, $V$ and $W$) of the electronic system. Details on the device fabrication and optical measurements are provided in Methods.

Figure 1c shows the reflectance contrast spectrum of device 1 as a function of electron density $\nu$ in the strong interaction limit ($U, V > W$). The three panels from left to right correspond to the moiré exciton of MoSe$_2$, the 2s exciton of the sensor layer, and the moiré exciton of WS$_2$, respectively. The moiré exciton spectra are consistent with a previous study [36]; the ground state exciton in WS$_2$ remains robust for the entire doping range, indicating that the electrons are doped only into the MoSe$_2$ layer in the type-I heterostructure. A series of incompressible states emerge at integer multiples and specific fractions of the moiré density. They modulate the moiré exciton features, but more significantly the sensor 2s exciton. At each incompressible state, the 2s exciton shows enhanced reflectance contrast or spectral weight (as well as spectral blueshift); it is consistent with small $\varepsilon$. The 2s exciton cannot be identified for the compressible states; it is merged into the band-to-band transitions [16]. The quenching of the 2s exciton is consistent with large negative dielectric constant for a metallic phase.

The insulating states at even integers ($\nu = 2$ and 4) are the single-particle moiré band insulators. The odd integer states ($\nu = 1$ and 3) are the Mott or charge-transfer insulators [32]. The fractional states (e.g. $\nu = 4/3, 3/2, 5/3$ etc.) are the Mott-Wigner insulators. Similar results are reported for a related WSe$_2$/WS$_2$ moiré superlattice [15, 26, 27, 30]. Generally, with increasing doping density the 2s exciton reflectance contrast decreases and less fractional states can be identified. It reflects the decreasing importance of the Coulomb interaction at large doping densities; the second moiré band has a larger bandwidth compared to the first moiré band (Methods). We study the second moiré band with doping density $\nu = 2 - 4$, for which the field-tuned MITs are more easily achieved. For the same field range, a weak electric-field effect is observed for the first moiré band with $\nu = 0 - 2$ (Extended Data Fig. 1).

Figure 2 illustrates the evolution of the incompressible states probed by the sensor 2s exciton for $\nu = 2 - 4$ under increasing electric fields. To enhance the optical contrast of these states, we show the energy derivative of the reflectance contrast spectrum $dR/d\epsilon$. As electric field increases, the insulating states gradually dissolve, first the band insulating state at $\nu = 4$, followed by the Mott-Wigner state at $\nu = 7/3$ and $8/3$, and the Mott state at $\nu = 3$. The band insulating state at $\nu = 2$ remains robust. The order of disappearance of these states can differ slightly in different devices. The result for device 2 (Extended Data Fig. 2) shows that the fractional states disappear first, followed by the $\nu = 4$ and 3 states; the latter two disappear at similar electric fields.

The above observation is consistent with the bandwidth-tuned MITs. As electric field increases (inducing larger $\Delta_c$ and shallower moiré potential), the moiré bandwidth $W$ increases; this is the predominant effect on the system parameters since $W$ is exponentially dependent on the moiré potential depth [4]. The disappearance of the $\nu = 4$ state indicates closing of the band gap between the second and third moiré band as $W$ increases. The vanishing $\nu = 3$ and the fractional filling states reflect closing of the Mott charge gap and the Mott-Wigner charge gap when $W$ becomes comparable to $U$ and $V$, respectively. Because $U > V$, the fractional states disappear at smaller critical fields than the $\nu = 3$ state in all devices examined in this study. On the other hand, the relative importance of the Mott gap and the band gap is sample dependent. A plausible origin is the twist angle and moiré density variations since these energies are generally charge-density dependent due to the strong correlation effects.

Next we investigate the bandwidth-tuned Mott and Wigner transitions systematically at fixed electron density of $\nu = 3$ and $7/3$, respectively. Figure 3a,b illustrate the sensor 2s exciton spectrum as a function of electric field. The 2s exciton resonance vanishes above a critical field, $E_c$. To determine the exciton spectral weight, we first normalize the reflectance contrast spectrum by that at a large field (e.g. 0.28 V/nm, above the critical field for $\nu = 3$). In the metallic phase, the 2s exciton resonance is quenched (Fig. 2); the reflectance contrast spectrum is dominated by a broad hump corresponding to the band-to-band transitions [16]; the spectrum is nearly identical to that at incommensurate fillings. Figure 3c shows the normalized spectrum at representative fields for $\nu = 3$ (the result for $\nu = 7/3$ is included in Extended Data Fig. 3). The integrated spectral weight (corresponding to the shaded area) is shown in Fig. 3d for $\nu = 3$ and $7/3$ as a function of electric field. As electric field or bandwidth increases, the spectral weight continuously decreases to zero. In addition, we do not observe any electric-field hysteresis within the experimental uncertainty. (The $\nu = 3$ spectral weight decreases for negative electric fields because the MoSe$_2$ moiré bands are approaching the WS$_2$ bands.)

We infer the dielectric constant of the moiré heterobilayer from the measured sensor exciton spectral weight by modeling excitons in the 2D sensor layer using realistic device geometry (Extended Data Fig. 4). We numerically solve the electron-hole Schrodinger equation with screened Coulomb potential by the heterobilayer and the hBN substrate with dielectric constant $\varepsilon$ and $\varepsilon_{BN}$, respectively. For $\varepsilon \gg \varepsilon_{BN}$, which holds for the insulating side near the transition, we obtain an empirical relation, $S \propto \varepsilon^{-0.7}$ (Methods). The field-dependence of $\varepsilon$ inferred using the relation is included in Fig. 3d (black lines). Here $\varepsilon$ is normalized by its value deep into the insulating phase, for which the 2s spectral weight plateaus. We also limit the electric-field range such that the signal-to-noise ratio of $S$ stays above 1. We find that $\varepsilon$ increases sharply towards the critical point for both $\nu = 3$ and $7/3$; the electric-field dependence of $\varepsilon$ is compatible with a power-law dependence, $\varepsilon \propto |E - E_c|^{-\gamma}$, with exponent $\gamma = 1.6 – 2.2$ for $\nu = 3$ and $1.2 – 1.8$ for $\nu = 7/3$ (Ext. Data Fig. 5).

The dielectric constant for a continuous Mott/Wigner transition is expected to diverge when the transition is approached from the insulating side [17-20], as polarization fluctuations and the holon/doublon density are proliferating near the critical electric field.

This is called the 'dielectric catastrophe' by Mott [17]. The dielectric constant is inversely proportional to the square of the insulator charge gap: $\varepsilon \propto \Delta^{-2}$ (Ref. [19]); the diverging $\varepsilon$ reflects a continuously vanishing $\Delta$. Our experiment thus supports a continuous Mott transition at $\nu = 3$ and a continuous Wigner transition at $\nu = 7/3$. The former also agrees with recent transport studies of other TMD moiré materials [5, 6]. However, a more quantitative analysis of the critical behavior in our experiment is not possible; the critical field and the exponent (that may reveal the universality class of the problem) cannot be accurately determined.

Finally, we examine the temperature dependence of these transitions. Figure 4a and 4b show the 2s exciton spectral weight as a function of electric field and temperature at $\nu = 3$ and 7/3, respectively. The spectral weight is always negligible on the metallic side; it gradually decreases on the insulating side with increasing temperature because the thermally excited free carriers in the moiré heterobilayer screens the excitonic interaction in the sensor. The melting temperature is respectively estimated to be ~ 65 K and 25 K for the $\nu = 3$ and 7/3 states. We compare the electric-field dependence of $S$ at 3.6 K and an elevated temperature for the two states in Fig. 4c and 4d. The spectral weight deep into the insulating phase is normalized to unity. At 3.6 K, the thermal excitation energy is small compared to the charge gap of both the Mott and Mott-Wigner insulators. Compared to the low-temperature behavior at 3.6 K, the MIT becomes a broadened metal-insulator crossover at high temperatures, a manifestation of critical point at lower temperatures. Our result thus suggests either a continuous Mott and Wigner transition with quantum critical point at zero temperature or weakly first-order transitions with critical point substantially below 3.6 K. The reduced dimensionality, the geometrically frustrated triangular lattice and the presence of disorders are known to favor continuous or weakly first-order transitions [9, 14, 21]. Above the critical point, these two scenarios are almost identical [37]; future experiments down to lower temperatures are required to distinguish them.

In conclusion, we have demonstrated bandwidth-tuned Mott and Wigner transitions at fixed band fillings of a Hubbard system based on semiconducting moiré materials. The transitions manifest a dielectric catastrophe when the critical point is approached from the insulating side. Our results present new opportunities to simulate Hubbard physics in the interesting regime of comparable Coulomb repulsion ($U, V$) and bandwidth ($W$), and to search for quantum spin liquids near the transitions [9, 12, 14, 21, 22].

**Methods**
**Device fabrication.** We fabricate dual-gate devices of a MoSe$_2$/WS$_2$ moiré heterobilayer with an integrated WSe$_2$ monolayer sensor using the reported dry transfer method [38]. Briefly, atomically thin flakes are first exfoliated from bulk crystals onto Si substrates and then stacked using a polymer stamp to form the desired heterostructure. Monolayer MoSe$_2$ and WS$_2$ flakes are angle aligned with a precision of about 0.5°. The orientation and relative alignment of these crystals are determined from the angle-resolved optical second harmonic measurement [30]. The WSe$_2$ sensor is separated from the moiré heterobilayer by a bilayer hBN. The TMD moiré heterobilayer and the sensor are grounded through few-layer graphite electrodes. The entire heterostructure is gated by hBN gate dielectrics (≈ 25 nm) and few-layer graphite gates on both sides.

**Optical reflectance contrast measurements.** Details of the reflectance contrast measurement are reported in the literature [15, 30]. Briefly, broadband white light from a tungsten-halogen lamp is focused under normal incidence to a diffraction-limited spot on the device by a high-numerical-aperture objective. The device is mounted in a closed-cycle cryostat with base temperature of 3.6 K (attoDry 1000). The reflected light is collected by the same objective and detected by a spectrometer with a liquid-nitrogen cooled charge-coupled device (CCD). The reflectance contrast spectrum $R \equiv (I' - I)/I$ is obtained by comparing the reflected light spectrum from the sample ($I'$) with a featureless background spectrum ($I$).

**Band alignment of the MoSe$_2$/WS$_2$ heterobilayer.** We determine the band alignment of the MoSe$_2$/WS$_2$ heterobilayer by examining a sample with large twist angle to avoid the moiré effect for simplicity. The reflectance contrast spectrum is measured as a function of doping density under zero applied electric field. Extended Data Fig. 6a and b show the fundamental exciton resonance in monolayer MoSe$_2$ and WS$_2$, respectively. The neutral exciton feature turns into the charged exciton (or polaron) feature in MoSe$_2$ with both electron and hole doping; the optical response of the WS$_2$ layer remains largely unperturbed. Charges are therefore introduced only into the MoSe$_2$ layer upon both electron and hole doping; the MoSe$_2$/WS$_2$ heterobilayer has a type-I band alignment.

We determine the band offsets by measuring the electric-field ($E$) dependence of the reflectance contrast spectrum. We choose a fixed electron doping density (≈ $3.7 \times 10^{12}$ cm$^{-2}$); the chemical potential is slightly above the conduction band edge of MoSe$_2$. As $E$ increases in the WS$_2$ to MoSe$_2$ direction ($E < 0$), the charged exciton feature in MoSe$_2$ changes to the neutral exciton above $E_0 \approx -0.33$ V/nm (Extended Data Fig. 6c); at the same time, the neutral exciton feature in WS$_2$ turns into the charged exciton (Extended Data Fig. 6d). The spectral changes correspond to the onset of charge transfer from MoSe$_2$ to WS$_2$ when the two conduction bands become nearly degenerate. Using the reported interlayer dipole moment in the MoSe$_2$/WS$_2$ heterobilayer [36], $d \approx 0.3$ e · nm, we estimate the conduction band offset to be $\Delta_C = d \cdot E_0 \approx 0.1$ eV. The valence band offset can be evaluated as $\Delta_V \approx E_g^W - E_g^{Mo} - \Delta_C \approx 0.32$ eV, where $E_g^W \approx 2.04$ eV and $E_g^{Mo} \approx 1.62$ eV are the optical gaps of monolayer MoSe$_2$ and WS$_2$, respectively. Extended Data Fig. 6e illustrates the inferred band alignment.

**Estimate of the bandwidth.** We estimate the first moiré conduction bandwidth, $W_0 \sim \frac{\hbar^2}{m a_M^2} \sim$ 5 - 10 meV, from the moiré period $a_M$. Here $\hbar$ and $m$ denote the Planck's constant and the conduction band mass of monolayer MoSe$_2$, respectively ($m \approx 0.56\, m_0$ in terms of the free electron mass $m_0$) [39]. The combined width of the first two moiré bands is $4W_0$; the second moiré bandwidth is about $3W_0$. These values are substantially smaller than $\Delta_C \approx 0.1$ eV. We focus on the second moiré conduction band with larger bandwidth; the bandwidth-tuned MITs are easier to achieve. The electrons reside in the MoSe$_2$ layer for all electric fields and doping densities in this study.

**Modeling the 2s exciton of the sensor layer.** Quantitative estimate of the effective dielectric constant of the moiré heterobilayer is obtained by modeling the sensor 2s exciton (Extended Data Fig. 4). We solve the Schrödinger equation, $H\Psi_{ns}(\rho) = E_{ns}\Psi_{ns}(\rho)$, for the energy ($E_{ns}$) and wavefunction ($\Psi_{ns}(\rho)$) of the $ns$ ($n = 1, 2, \ldots$) exciton state using the finite difference method [40]. For radially symmetric $ns$ excitons confined in the 2D sensor plane, the Hamiltonian is given by $H = -\frac{\hbar^2}{2m_R}\left(\partial_\rho^2 + \frac{1}{\rho}\partial_\rho\right) + V(\rho)$, where $\rho$ is the distance between the electron and hole, $m_R$ ($\approx 0.2\, m_0$ [Ref. [41]]) is the reduced mass of the exciton, and $V(\rho)$ is the electrostatic potential between the electron and hole. We model $V(\rho)$ using device geometry shown in Extended Data Fig. 4a. The thickness of the sensor layer is ignored; the thickness of the hBN spacer and the moiré heterobilayer are $d_1$ and $d_2$, respectively; the hBN gate dielectric is assumed to be infinitely thick. The latter is a good approximation when the exciton Bohr radius does not exceed substantially the hBN thickness and screening by the graphite gates is negligible. We express the potential as follows [42]

$$V(\rho) = -\frac{e^2}{4\pi\varepsilon_0\varepsilon_{BN}}\left[-\frac{1}{\rho} + \sum_{j=0}^{\infty}\beta^{2j+1}\left(\frac{1}{\sqrt{\rho^2+4(d_1+jd_2)^2}} - \frac{1}{\sqrt{\rho^2+4(d_1+(j+1)d_2)^2}}\right)\right]. \quad (1)$$

Here $\varepsilon_0$ is the vacuum permittivity; $\beta = (\varepsilon - \varepsilon_{BN})/(\varepsilon + \varepsilon_{BN})$ is given by the dielectric constant of hBN ($\varepsilon_{BN} = 4.5$ [Ref. [15]]) and the heterobilayer ($\varepsilon$); we take $d_1 = 0.9$ nm for a 2L-hBN spacer and $d_2 = 0.6$ nm for electrons residing in the MoSe$_2$ layer of the moiré. We assume $\varepsilon$ to be a real value for the insulating states for simplicity.

Extended Data Fig. 4b and 4c illustrate the spatial distribution of potential $V(\rho)$ and wavefunction $\Psi_{2s}(\rho)$ of the 2s exciton for several values of $\varepsilon/\varepsilon_{BN}$. The 2s exciton radius $r_{2s}$ ($=\sqrt{\langle\Psi_{2s}|\rho^2|\Psi_{2s}\rangle}$) and binding energy $E_{2s}$ as a function of $\varepsilon/\varepsilon_{BN}$ are shown in Extended Data Fig. 4d and 4e, respectively. For $\varepsilon/\varepsilon_{BN} = 2$, screening by the heterobilayer is negligible; we have $V(\rho) \approx -\frac{e^2}{4\pi\varepsilon_0\varepsilon_B\rho}$ and $r_{2s} \approx 4.7$ nm. The latter agrees well with the reported value of $r_{2s} \approx 6.6$ nm for monolayer WSe$_2$ embedded in hBN [40]. As $\varepsilon/\varepsilon_{BN}$ increases, $V(\rho)$ is suppressed and $\Psi_{2s}(\rho)$ is flattened; $r_{2s}$ increases and $E_{2s}$ decreases. For $\varepsilon/\varepsilon_{BN} = 10^4$, $r_{2s}$ exceeds 50 nm. We limit the range of $\varepsilon/\varepsilon_{BN}$ to $< 10^4$ (so that the correction from the gate screening effect remains small) and perform a power-law analysis of $E_{2s}$. The binding energy is well described by $\sim \left(\frac{\varepsilon}{\varepsilon_{BN}}\right)^{-0.7}$ for $\frac{\varepsilon}{\varepsilon_{BN}} > 10$ (solid line in Extended Data Fig. 4e). The exciton spectral weight is expected to follow the same scaling law on the dielectric constant since both the exciton spectral weight and binding energy scale quadratically with the exciton radius [43] (Extended Data Fig. 4f shows $E_{2s} \sim (r_{2s})^{-2}$ for the entire range of dielectric constant). We examine the result for several hBN space thicknesses. The power-law exponent of $-0.7$ remains a good approximation as long as the spacer thickness is much smaller than the 2s exciton radius. In the main text, we use the power-law dependence to extract the evolution of the dielectric constant on the out-of-plane electric field from the experimental spectral weight.

## Figures

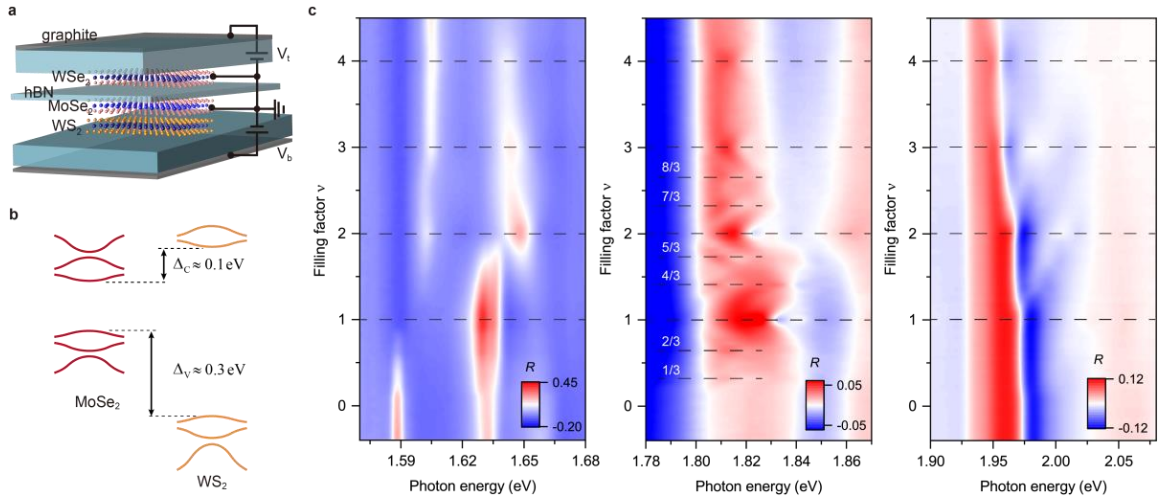

**Figure 1 | Correlated insulating states in MoSe$_2$/WS$_2$ moiré heterobilayers. a,** Schematic illustration of a dual-gated MoSe$_2$/WS$_2$ moiré heterobilayer with an integrated WSe$_2$ monolayer sensor separated by bilayer hBN. Voltage $V_t$ and $V_b$ are applied to the top and bottom hBN-graphite gates, respectively. Both the moiré heterobilayer and the sensor are grounded. **b,** Type-I band alignment in MoSe$_2$/WS$_2$ moiré heterobilayer with the conduction band offset $\Delta_c \approx 0.1$ eV and valence band offset $\Delta_v \approx 0.3$ eV. **c,** Reflectance contrast spectrum of device 1 as a function of electron doping density $\nu$ (in units of the moiré density). Full lattice fillings are denoted by dashed lines. The panels from left to right correspond to the moiré exciton of MoSe$_2$, the sensor 2s exciton, and the moiré exciton of WS$_2$, respectively. The incompressible states are identified by the enhanced reflectance contrast of the 2s exciton.

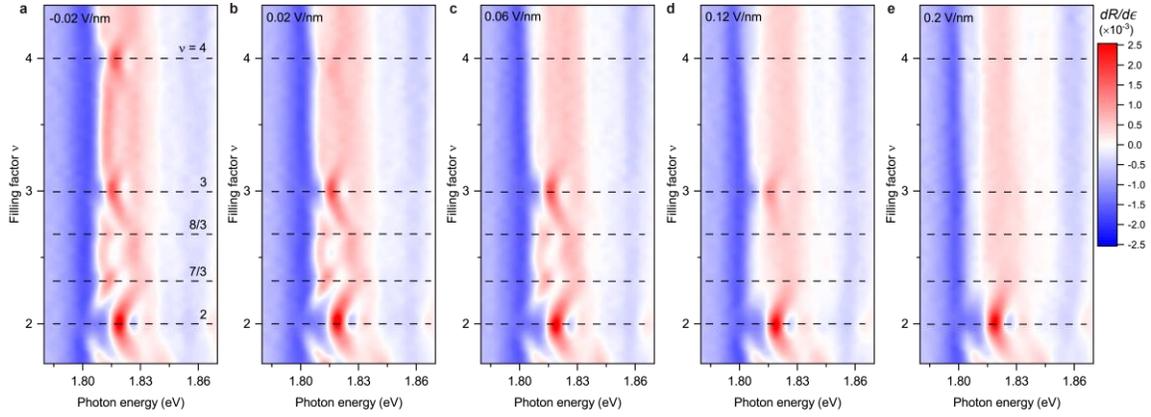

**Figure 2 | Electric-field tuned MITs. a-e,** The energy-derivative of the reflectance contrast spectrum ($dR/d\epsilon$) of the sensor 2s exciton as a function of electron filling factor ($\nu$ = 2 - 4) at electric field $E$ ranging from -0.02 V/nm to 0.2 V/nm. The incompressible states corresponding to the enhanced reflectance contrast at $E$ = -0.02 V/nm are labeled by the dashed black lines. They gradually dissolve as electric field increases.

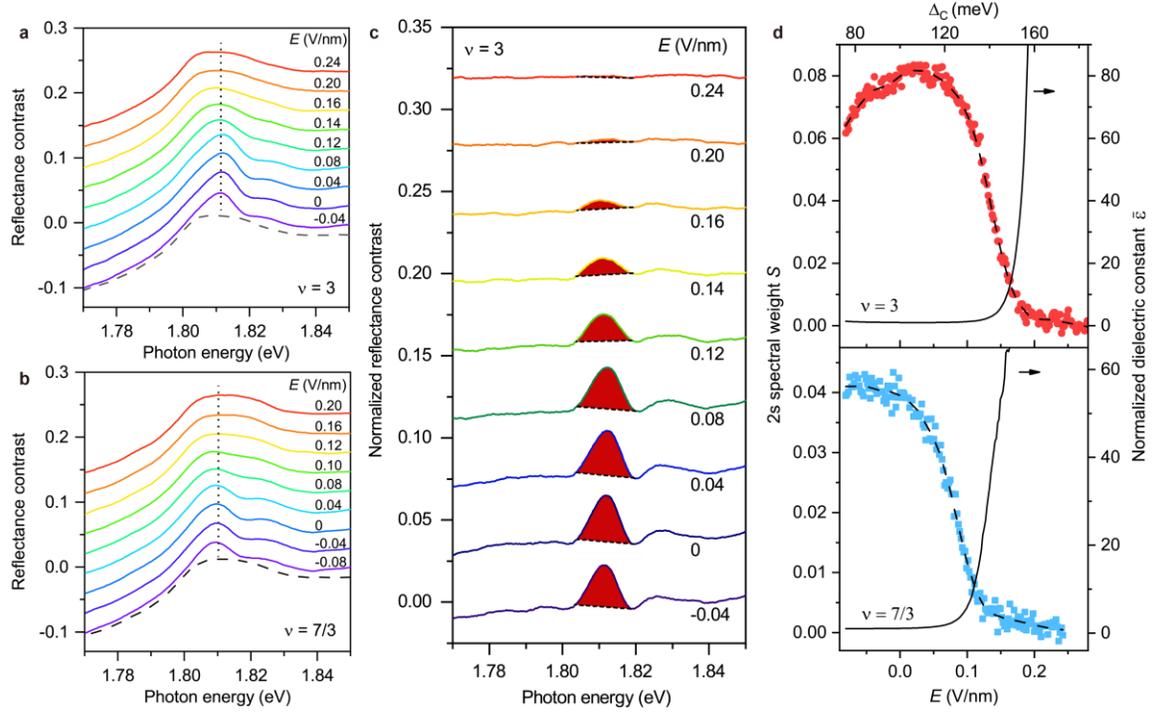

**Figure 3 | Dielectric catastrophe near the Mott and Wigner transitions. a, b,** Reflectance contrast of the sensor 2s exciton as a function of photon energy and electric field at fixed electron filling factor of $\nu = 3$ (**a**) and 7/3 (**b**). The spectra are vertically displaced by a constant 0.03. For comparison, the reflectance contrast spectra at high electric fields (0.28 V/nm and 0.24 V/nm for $\nu = 3$ and 7/3, respectively) are also shown as dashed lines. The vertical dotted lines denote the 2s exciton peak. The broad hump at high electric fields corresponds to the band-to-band transition. **c,** Reflectance contrast spectrum at $\nu = 3$, normalized by that at 0.28 V/nm (in the metallic phase), at several representative electric fields. The spectral weight $S$ is extracted by integrating the shaded area. **d,** The extracted 2s spectral weight $S$ (symbols) as a function of electric field (bottom axis) and $\Delta_c$ (top axis) for $\nu = 3$ (upper panel) and $\nu = 7/3$ (lower panel). The band offset $\Delta_c$ is calculated using the applied electric field as described in Methods. The dashed lines are the smoothed data using the Savitzky-Golay algorithm with a window of 80 mV/nm. The solid lines are the dielectric constant $\bar{\varepsilon}$ of the moiré heterobilayer (right axis) that is normalized to unity at 0.02 V/nm and -0.08 V/nm (upper and lower panel, deep in the insulating phase). It is obtained from the dashed lines using the empirical relation, $S \propto \varepsilon^{-0.7}$ (Methods).

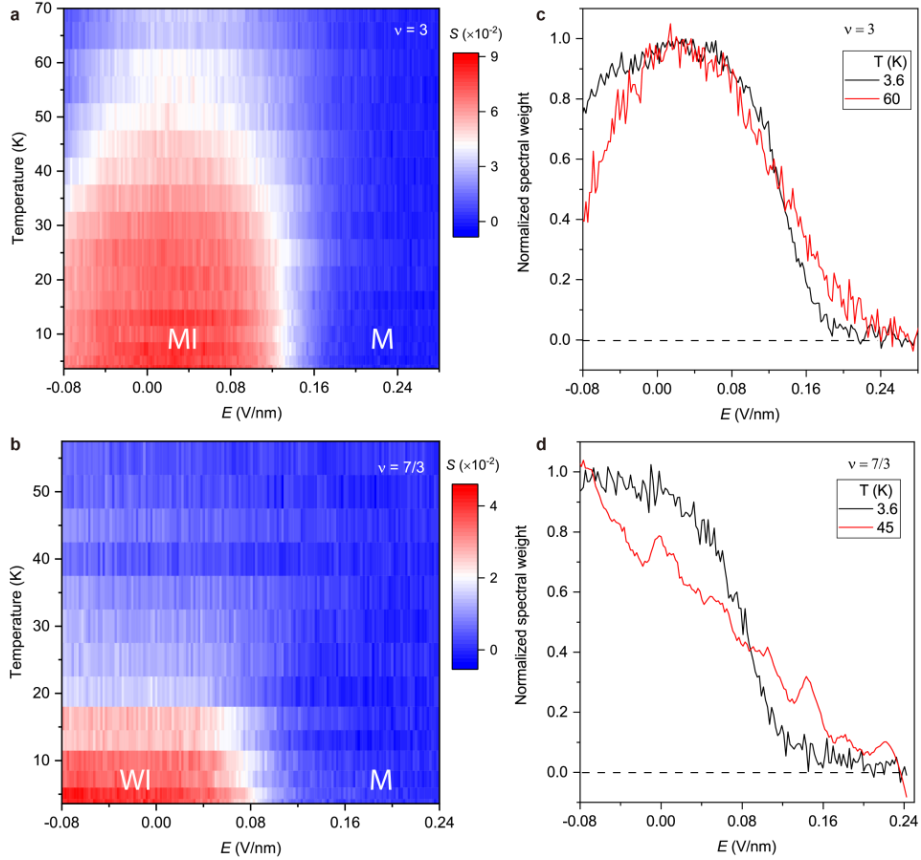

**Figure 4 | Temperature dependence. a, b,** Contour plot of the sensor 2s spectral weight as a function of temperature and electric field for $\nu = 3$ (**a**) and $\nu = 7/3$ (**b**). MI, WI, and M represent, respectively, the Mott insulator, Mott-Wigner insulator, and metal. Regions with enhanced spectral weight (red) correspond to the incompressible states (MI and WI); regions with negligible spectral weight (blue) corresponds to compressible states (metal). **c, d**, Electric-field dependence of the 2s spectral weight at selected temperatures for $\nu = 3$ (**c**) and $\nu = 7/3$ (**d**). The spectral weight maximum is normalized to unity. The metal-insulator crossover is broadened at high temperatures.

**Extended data figures**

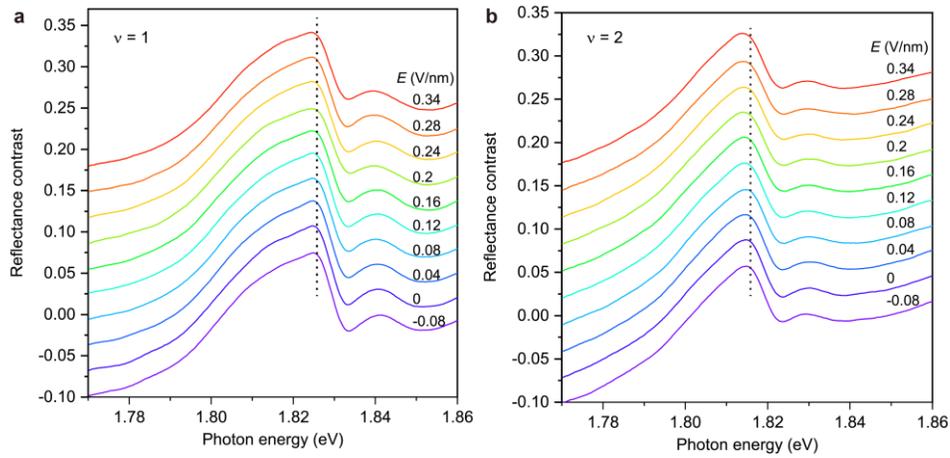

**Extended Data Figure 1 | Electric field dependence at $\nu = 1$ and $\nu = 2$.** The reflectance contrast spectrum near the sensor 2s exciton as a function of photon energy and electric field at $\nu = 1$ (**a**) and $\nu = 2$ (**b**). The spectra are vertically displaced by a constant 0.03 for clarity. The vertical dotted lines trace the 2s exciton peak. Negligible electric field dependence is observed.

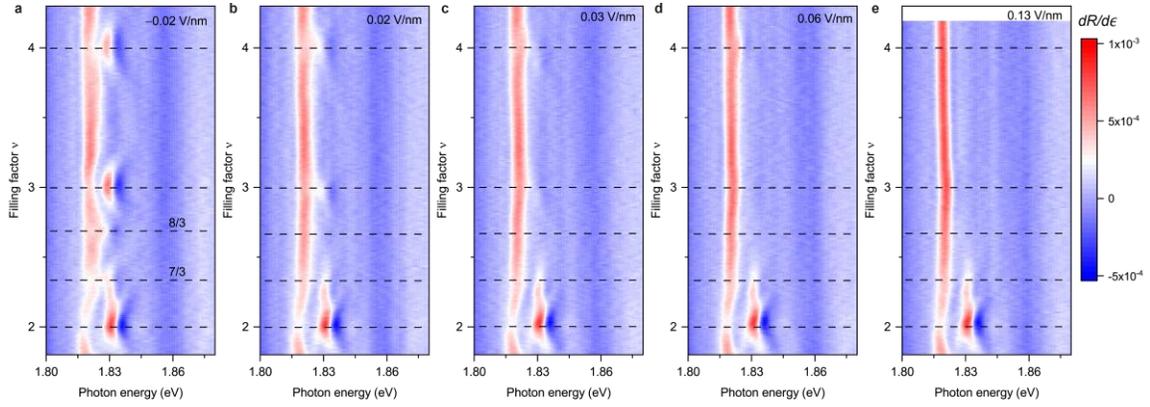

**Extended Data Figure 2 | Electric-field tuned MITs for device 2. a-e,** The energy-derivative of the reflectance contrast spectrum ($dR/d\epsilon$) of the sensor 2s exciton as a function of electron filling factor ($\nu$ = 2 - 4) at electric field $E$ ranging from -0.02 V/nm to 0.13 V/nm. The incompressible states corresponding to the enhanced reflectance contrast at $E$ = -0.02 V/nm are labeled by the dashed black lines. They gradually dissolve as electric field increases. The $\nu = 4$ and $\nu = 3$ insulating states disappear at a similar electric field in this device.

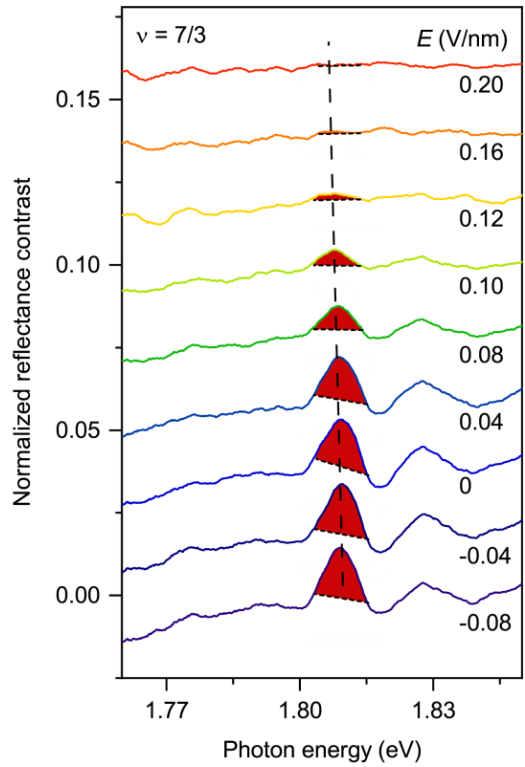

**Extended Data Figure 3 | Normalized reflectance contrast spectra as a function of electric field at $\nu = 7/3$.** The reflectance contrast spectra are normalized to that at 0.24 V/nm, where no incompressible state can be identified. The 2s exciton oscillator strength (shaded area) decreases to zero at large electric fields. The spectra are vertically displaced by a constant 0.02. The dashed line traces the 2s exciton peak.

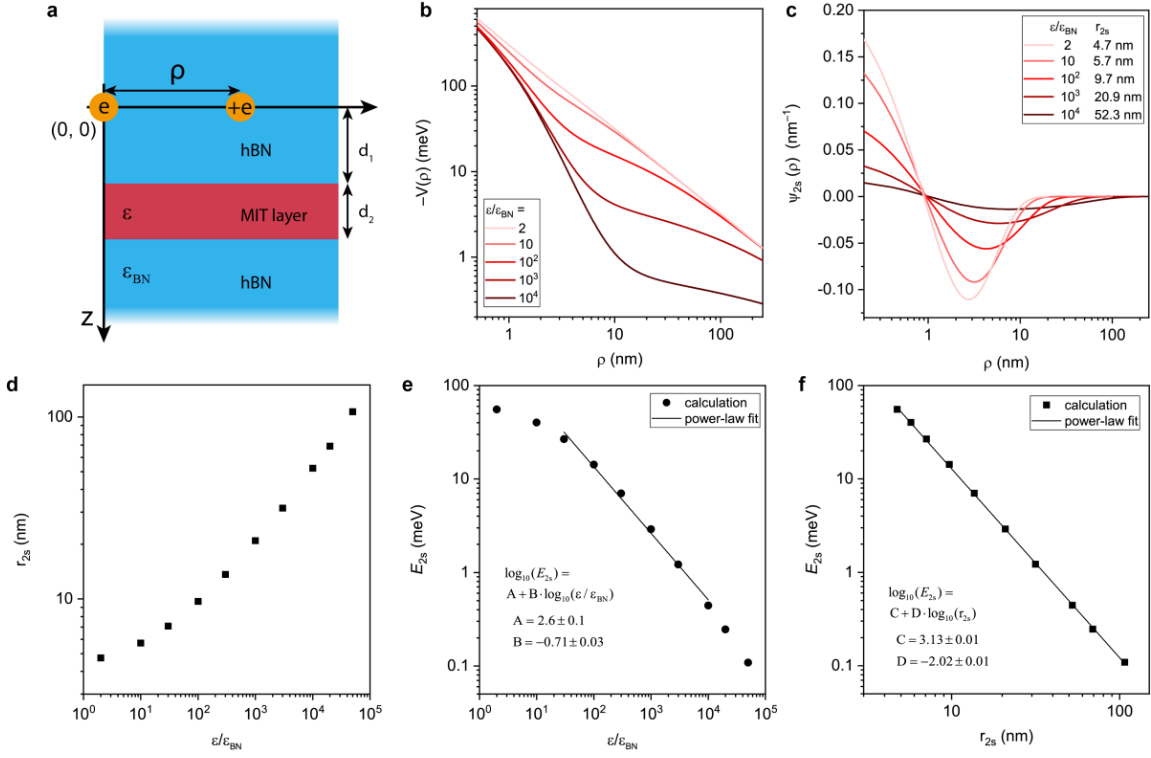

**Extended Data Figure 4 | Results for the 2s exciton Rydberg state. a,** Schematic structure of the device geometry minus the gates. An electron-hole pair with separation $\rho$ in the sensor layer is separated from the moiré sample (with thickness $d_2$) by a distance $d_1$. The sample and the sensor are both encapsulated by hBN (blue). $d_1$ and $d_2$ are set at 0.9 nm and 0.6 nm, respectively, to represent the realistic device geometry. **b**, **c**, Dependence of the screened electron-hole Coulomb potential $-V(\rho)$ (**b**) and the 2s exciton wavefunction $\Psi_{ns}(\rho)$ (**c**) on the electron-hole separation $\rho$ for varying ratio of the dielectric constants $\frac{\varepsilon}{\varepsilon_{BN}}$. **d, e**, Dependence of the 2s Bohr radius (**d**) and the 2s binding energy on the ratio $\frac{\varepsilon}{\varepsilon_{BN}}$. The black line is a power-law fit to the result for $\frac{\varepsilon}{\varepsilon_{BN}}$ between $10^2$ and $10^4$. **f**, Dependence of the 2s binding energy on the 2s Bohr radius. The black line is a power-law fit to the result.

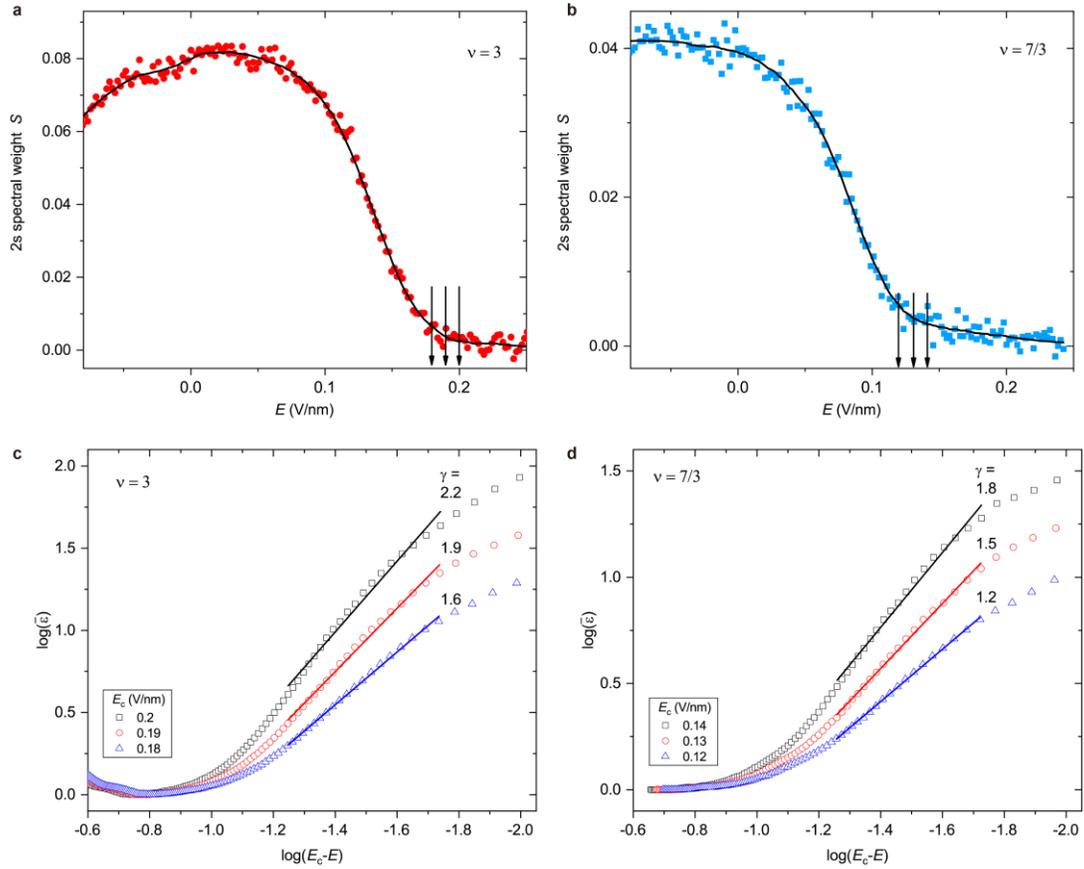

**Extended Data Figure 5 | Critical exponents for the MITs. a**, **b**, The 2s exciton spectral weight as a function of the electric field at $\nu = 3$ (**a**) and $7/3$ (**b**). The solid lines are the smoothed data using the Savitzky-Golay algorithm with a window of 80 mV/nm. The arrows denote the chosen critical electric fields $E_c$ of the MITs; the value of $E_c$ cannot be determined accurately in our experiment. **c**, **d**, Dependence of the normalized dielectric constant $\bar{\varepsilon}$ on the reduced electric field $E - E_c$ in log-log scale for $\nu = 3$ (**c**) and $7/3$ (**d**). The solid lines are linear fits to the data near $E_c$ over a limited range of $E - E_c$ in order to estimate the value of the critical exponent $\gamma$, which varies from 1.6 to 2.2 for $\nu = 3$ and from 1.2 to 1.8 for $\nu = 7/3$. Different colors correspond to different chosen values of $E_c$.

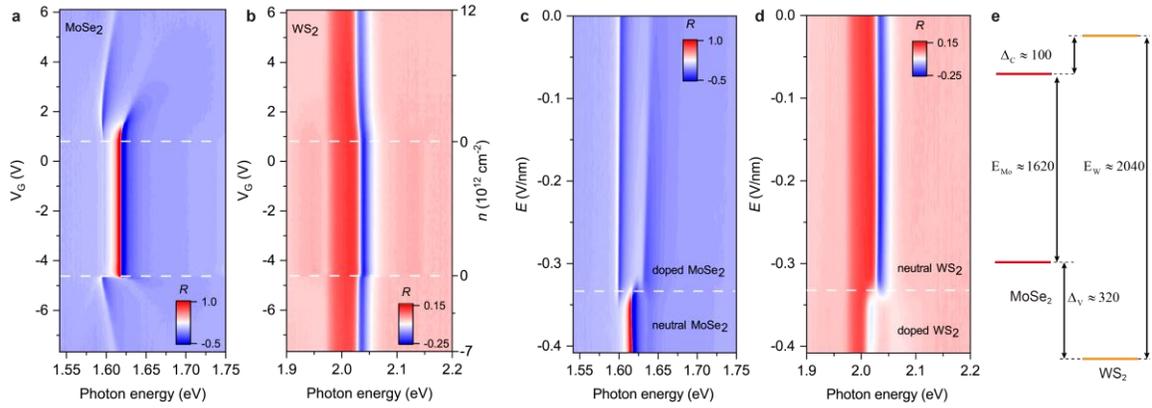

**Extended Data Figure 6 | Band alignment in large-twist-angle MoSe$_2$/WS$_2$. a**, **b**, The doping density dependent reflectance contrast spectrum of the MoSe$_2$ (**a**) and WS$_2$ (**b**) intralayer excitons under zero electric field. The gate voltage $V_G$ here only changes the doping density ($n$). The white dashed lines indicate the onsets of electron and hole doping. Charged excitons are only observed in the MoSe$_2$ layer, consistent with a type-I band alignment in MoSe$_2$/WS$_2$ heterobilayer. **c**,**d**, The electric field dependent reflectance contrast spectrum of the MoSe$_2$ (**c**) and WS$_2$ (**d**) intralayer excitons at a constant electron doping density $3.7\times10^{12}$ cm$^{-2}$. The white dashed lines, where the transition from charged exciton to neutral exciton occurs, correspond to the onset of charge transfer between MoSe$_2$ and WS$_2$. The electric field where this happens allows us to determine the band offsets. **e**, The determined band alignment in MoSe$_2$/WS$_2$ heterobilayer (Methods). Energies are denoted in meV. E$_{Mo}$ and E$_W$ are the optical gaps of MoSe$_2$ and WS$_2$, respectively. $\Delta_C$ and $\Delta_V$ are the conduction and valence band offsets, respectively.